\documentclass[aps,amsfonts,reprint,tightenlines,amssymb,superscriptaddress,twocolumn]{revtex4-2}  

\usepackage{graphicx}
\usepackage{dcolumn}
\usepackage{bm}
\usepackage{mathtools}
\usepackage{tabularx}

\usepackage{qcircuit}
\usepackage[caption=false]{subfig}

\usepackage{algorithm}
\usepackage[noend]{algpseudocode}
\algrenewcommand\algorithmicdo{}

\makeatletter
\renewcommand{\ALG@name}{Procedure}
\makeatother

\usepackage[linktocpage=true,
  colorlinks=true, 
  pdfborder={0 0 0},
  linkcolor=blue,
  citecolor=red,
  filecolor=yellow,
  urlcolor=blue,
  bookmarks,
  pdfauthor={},
]{hyperref}

\usepackage{ifthen}
\newcounter{is_qcircuit_used}
\usepackage{xcolor}
\usepackage{pagecolor}
\definecolor{offwhite}{RGB}{250, 245, 240}

\begin{document}

\preprint{APS/123-QED}
\title{Quadratic acceleration of multi-step probabilistic algorithms for state preparation}

\author{Hirofumi Nishi}
\email{nishi.h.ac@m.titech.ac.jp}
\affiliation{
Laboratory for Materials and Structures,
Institute of Innovative Research,
Tokyo Institute of Technology,
Yokohama 226-8503,
Japan
}
\affiliation{
Quemix Inc.,
Taiyo Life Nihombashi Building,
2-11-2,
Nihombashi Chuo-ku, 
Tokyo 103-0027,
Japan
}

\author{Taichi Kosugi}
\affiliation{
Laboratory for Materials and Structures,
Institute of Innovative Research,
Tokyo Institute of Technology,
Yokohama 226-8503,
Japan
}

\affiliation{
Quemix Inc.,
Taiyo Life Nihombashi Building,
2-11-2,
Nihombashi Chuo-ku, 
Tokyo 103-0027,
Japan
}

\author{Yusuke Nishiya}
\affiliation{
Laboratory for Materials and Structures,
Institute of Innovative Research,
Tokyo Institute of Technology,
Yokohama 226-8503,
Japan
}

\affiliation{
Quemix Inc.,
Taiyo Life Nihombashi Building,
2-11-2,
Nihombashi Chuo-ku, 
Tokyo 103-0027,
Japan
}

\author{Yu-ichiro Matsushita}
\affiliation{
Laboratory for Materials and Structures,
Institute of Innovative Research,
Tokyo Institute of Technology,
Yokohama 226-8503,
Japan
}
\affiliation{
Quemix Inc.,
Taiyo Life Nihombashi Building,
2-11-2,
Nihombashi Chuo-ku, 
Tokyo 103-0027,
Japan
}
\affiliation{
Quantum Material and Applications Research Center,
National Institutes for Quantum Science and Technology,
2-12-1, Ookayama, Meguro-ku, Tokyo 152-8552, Japan
}

\date{\today}
\begin{abstract}
For quantum state preparation, a non-unitary operator is typically designed to decay undesirable states contained in an initial state using ancilla qubits and a probabilistic action. 
Probabilistic algorithms do not accelerate the computational process compared to classical ones. 
In this study, quantum amplitude amplification (QAA) and multi-step probabilistic algorithms are combined to achieve quadratic acceleration. 
This method outperforms quantum phase estimation in terms of infidelity.
The quadratic acceleration was confirmed by the probabilistic imaginary-time evolution (PITE) method.
\end{abstract}

\maketitle

\paragraph*{Introduction.} ---
The accurate and efficient calculation of ground states is of considerable importance in the field of quantum physics because it provides key insights into the properties and behavior of diverse quantum systems.
Quantum phase estimation (QPE) \cite{Kitaev1995arXiv, Abrams1999PRL, Ding2023arXiv, Ding2023PRXQ} is a promising quantum algorithm that estimates the eigenvalues of input eigenvectors of a Hamiltonian more efficiently than classical computers. 
Despite the advantages of QPE, the preparation of an input state as close as possible to the ground state remains a problem. 
If the input state of the QPE contains only a small portion of the ground state, the probability of obtaining the ground-state energy decreases, as indicated by the scaling $\mathcal{O}(|c_1|^2)$, where $|c_1|^2$ denotes the weight of the ground state in the input state.
Thus, significant research has been conducted on quantum algorithms for ground-state preparation \cite{Kadowaki1998PRE, Farhi2000arXiv, AspuruGuzik2005Science, Poulin2009PRL, Ge2019JMP, Lin2020Quantum, Lin2022PRXQ, Choi2021PRL, Silva2021arXiv, Kosugi2022PRR, Meister2022arXiv, Stetcu2022arXiv, Xie2022arXiv, Chan2023arXiv}.

State-preparation schemes based on non-unitary operations have been proposed to obtain the ground state \cite{Choi2021PRL, Silva2021arXiv, Kosugi2022PRR, Meister2022arXiv, Stetcu2022arXiv, Xie2022arXiv, Chan2023arXiv}.
Previous studies have realized imaginary-time evolution (ITE) operators \cite{Silva2021arXiv, Kosugi2022PRR, Xie2022arXiv, Chan2023arXiv}, cosine functions \cite{Choi2021PRL, Meister2022arXiv}, and shifted step functions \cite{Lin2020Quantum, Lin2022PRXQ} on quantum computers using ancilla qubits and a probabilistic method, both of which rely on forward- and backward-controlled real-time evolution (CRTE) operators. 
The implementation of real-time evolution (RTE) operators on quantum computers has been well established in the context of Hamiltonian simulations based on the Trotter decomposition \cite{ Seth1999Science, Abrams1997PRL, Kassal2008PNAS, Childs2019PRL, Childs2021PRX}, Taylor series \cite{Childs2012QIC, Berry2015PRL}, and qubitization \cite{Low2017PRL, Low2019Quantum, Gilyen2019ACM, Martyn2021PRXQuantum}. 
Thus, any sophisticated implementation of RTE operators can be incorporated into state-preparation schemes using non-unitary quantum circuits.

The computational costs of quantum algorithms that implement non-unitary operators probabilistically have also been estimated for the cosine function \cite{Ge2019JMP, Choi2021PRL, Meister2022arXiv} and ITE operators \cite{Nishi2023arXiv}. 
These quantum algorithms incur $\mathcal{O}(|c_1|^{-2})$ computational costs to obtain the ground state. 
In particular, if $|c_1| = 1/\sqrt{N}$, where $N=2^n$ and $n$ denotes the number of qubits, i.e., in a scenario where even the approximate ground state is not known, at least $\mathcal{O}(N)$ computational cost is incurred. This implies that quantum acceleration is not realized.
The complexity class of ground-state preparation is known to be Quantum Merlin-Arthur \cite{Kitaev2002Book, Kempe2005Book, Oliveira2008arXiv}. 
In this study, we propose a multi-step quantum algorithm that achieves quadratic acceleration of the probabilistic state preparation scheme.
The proposed quantum algorithm utilizes quantum amplitude amplification (QAA), which enhances the probability of obtaining a desired state based on repeated operations.
First, we summarize quantum algorithms for probabilistic formalisms to implement non-unitary operators and estimate their computational costs. We highlight that existing probabilistic algorithms have a computational scaling of order $\mathcal{O}(|c_1|^{-2})$ for state preparation, which compromises quantum advantages. 
Subsequently, we propose quantum algorithms to achieve quadratic acceleration.

\begin{figure}[ht]
        \centering
        \includegraphics[width=0.42 \textwidth]{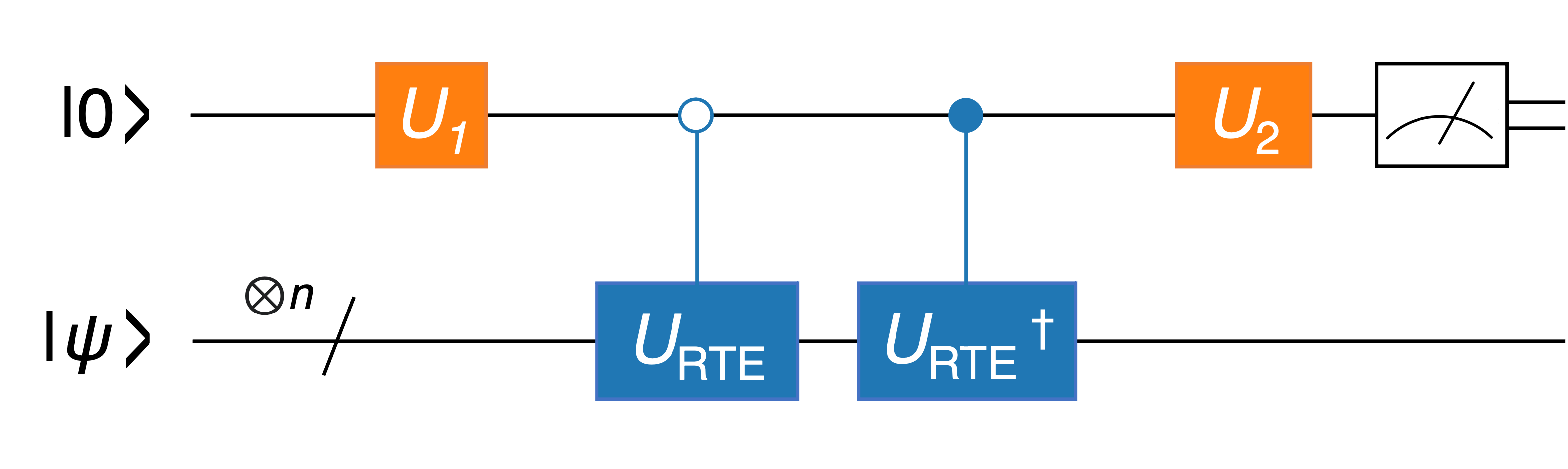}
 \caption{
Quantum circuit of probabilistic algorithm for ground state preparation. The circuit is composed of a single ancilla qubit and forward- and backward CRTE gates. 
}
 \label{fig:qc_pite}
\end{figure}

\paragraph*{Ground-state preparation.}---
Let us consider a nonunitary operator $f(\mathcal{H})$, where $\mathcal{H}$ is an $n$-qubit system Hamiltonian. 
$f(\mathcal{H})$ is embedded in the extended unitary matrix by introducing an ancilla qubit as follows:
\begin{gather}
    \mathcal{U}
    \equiv
    \begin{pmatrix}
        f(\mathcal{H})  & \sqrt{1-f^{2}(\mathcal{H})} \\
        \sqrt{1-f^{2}(\mathcal{H})} & - f(\mathcal{H})  
    \end{pmatrix}
    ,
\end{gather}
comprising submatrices coupled with the ancillary $| 0 \rangle$ and $| 1 \rangle$ states.
The desired state is obtained when the ancilla is $| 0 \rangle,$
whereas a state coupled to $| 1 \rangle$ is undesirable.
In particular, the action of the unitary matrix $\mathcal{U}$ on the input state leads to
\begin{gather}
    f(\mathcal{H}) |\psi\rangle \otimes |0\rangle
    +
    \sqrt{1 - f^{2}(\mathcal{H})}|\psi\rangle \otimes |1\rangle .
\end{gather}
Any unitary matrix $\mathcal{U}$ can be decomposed into at least the first order of $\mathcal{H}$ (see details in Supplementary Materials (SM) \cite{SM}) and implemented using forward- and backward CRTE operations and single-qubit gates (Fig. \ref{fig:qc_pite}). 

During repeated operation, $\mathcal{U}$ involves certain parameters, such as the real time-step size $\Delta\tau$ depicted in Fig. \ref{fig:qc_pite}, which can be selected freely at each step. 
The unitary matrix $\mathcal{U}$ at the $k$-th step is denoted by $\mathcal{U}_{k}$. 
When $K$ ancilla qubits are used, with a different one used for each $\mathcal{U}_{k}$ without measuring them, the actions of $\{\mathcal{U}_{k}\}$ on the input state with the initialized ancilla qubits yield
\begin{gather}
    F_{K}(\mathcal{H})
    |\psi\rangle \otimes |0\rangle^{\otimes K}
    +
    \mathrm{(other~states)} .
\label{eq:state_after_pite_Kstep}
\end{gather}
where
$
    F_{K}(\mathcal{H})
    \equiv
    \prod_{k=1}^{K} f_{k}(\mathcal{H})
$.
The use of only one ancilla qubit is permitted if we reuse it based on the previous step after measuring and initializing it. 
In \cite{Kosugi2022PRR} and \cite{Choi2021PRL}, $f_{k}(\mathcal{H})$ was used as an approximate ITE operator $e^{-\Delta \tau_k \mathcal{H}}$ and a cosine function $\cos(t_k \mathcal{H})$, respectively. 
Specific quantum circuits are summarized in SM \cite{SM}.
Repeated operations of $f_{k} (\mathcal{H})$ decay undesirable states, such as excited states. 
Quantum signal processing is another approach to design such decay functions \cite{Low2019Quantum}, where the shifted step function is approximated using a Chebyshev polynomial and implemented using a single ancilla qubit \cite{Lin2020Quantum, Lin2022PRXQ}.

Observing all ancilla qubits in $|0\rangle$ state leads to the collapse of the entangled wave function to 
\begin{gather}
    |\Psi_{K} \rangle
    =
    \frac{1}{\sqrt{P_{K}}} F_{K}(\mathcal{H}) |\psi\rangle ,
\end{gather}
where $P_{K}$ denotes the total probability of all steps being successful.
The input state is expanded as follows:
$|\psi\rangle = \sum_{i=1}^{N} c_{i}|\lambda_i \rangle$ 
where $| \lambda_i \rangle$ denotes the $i$th eigenstate of the Hamiltonian $\mathcal{H}$ and $c_i$ denotes the expansion coefficient. 
For simplicity, we assume a non-degenerate and ascending order of eigenvalues; however, the generalization is straightforward.
If $F_{K}(\mathcal{H})$ is well-designed to decay the undesired state, then the total success probability becomes
\begin{gather}
    P_{K}
    =
    \frac{1}{1-\delta_K}|c_{1}|^{2} ,
\end{gather}
where $\delta_K$ denotes the infidelity, defined as $\delta_K \equiv 1 - \mathcal{F}_K$, with fidelity $\mathcal{F}_K \equiv | \langle \lambda_{1}|\Psi_{K}\rangle |^{2}$ (for further details, see SM \cite{SM}). 
Here, we assume $F_K(\lambda_1)=1$, which is realized by a constant energy shift in the PITE \cite{Nishi2023arXiv}.
Every quantum algorithm that decays the undesirable state to achieve a small value of $\delta_K$ using a non-unitary operation exhibits $\mathcal{O}(|c_{1}|^{2})$ scaling of the total success probability. 
Importantly, this implies that the scaling of the total success probability is independent of the type of nonunitary operator $f_{k}(\mathcal{H})$, the number of ancilla qubits, or the circuit implementation method for $f_{k}(\mathcal{H})$. 
If the approximate ground state is not known, e.g., $|c_{i}|^{2} = 1/N$ for each $i$, the computational cost of at least one success scale is of the order $\mathcal{O}(N)$. 
No quantum acceleration is observed in these algorithms.
In contrast, CRTE gates are efficiently implemented at polynomial cost \cite{Seth1999Science, Abrams1997PRL, Kassal2008PNAS, Childs2019PRL, Childs2021PRX, Childs2012QIC, Berry2015PRL, Low2017PRL, Low2019Quantum, Gilyen2019ACM, Martyn2021PRXQuantum}.
For example, the circuit depth for CRTE for an $n_e$-electron system based on the first quantization Hamiltonian scales with the order of 
$
    d_{\mathrm{CRTE}}
    =
    \mathcal{O}(
        r n_e^2 \operatorname{poly}(
            \log (n_e^{1 / 3} /\Delta x)
        )
    )
    ,
$
where $\Delta x$ represents the grid spacing of the discretized space, and $r$ denotes the Trotter number dividing the imaginary time-step size 
\cite{Kassal2008PNAS, Kosugi2022PRR}.
Then, the computational cost is given by 
\begin{gather}
    \frac{d_{\mathrm{CRTE}} K}{P_{K}} 
    =
    \mathcal{O} \left(
        \frac{d_{\mathrm{CRTE}}}{|c_{1}|^{2}}
        \ln \left(
            \frac{(1-\delta_K)(1-|c_1|^{2})}{\delta_K |c_1|^2}
        \right) 
    \right) ,
\label{eq:computational_cost_pite}
\end{gather}
(For details, see SM \cite{SM}).

\begin{figure*}[ht]
        \centering
        \includegraphics[width=0.90 \textwidth]{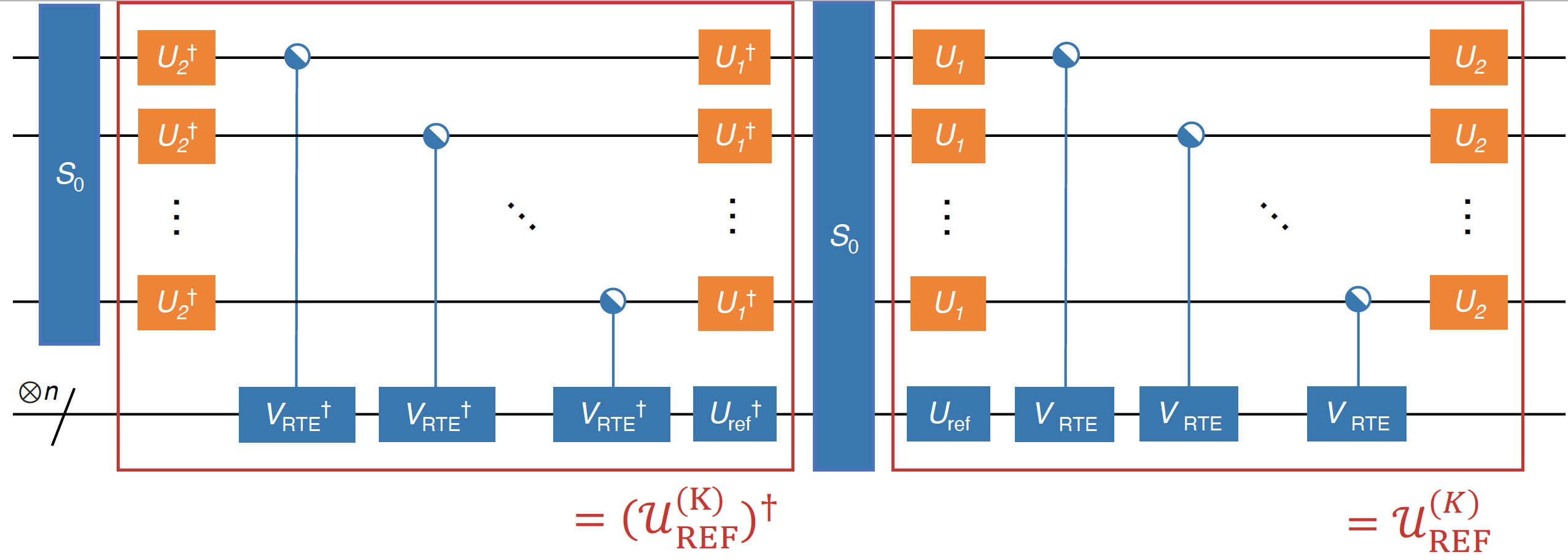}
 \caption{
Quantum circuit for the amplitude amplification operator of the probabilistic algorithm for ground-state preparation given by Eq. (\ref{eq:amplitude_amplification_opr}) while ignoring the global phase. The forward and backward CRTE gates are combined and represented by the controlled-$V_{\mathrm{RTE}}$ gate.
On the left, $S_0$ represents $S_0^{(K)}(\phi_{2i}),$ while that on the right represents $S_0^{(n+K)}(\phi_{2i-1})$.
$U_{\mathrm{ref}}$ is defined as $U_{\mathrm{ref}}|0\rangle^{\otimes n} \equiv |\psi\rangle$, which is the input state before the PITE circuit. 
In the red box on the right, $V_{\mathrm{RTE}} = V_{\mathrm{RTE}}(t_1)$, $V_{\mathrm{RTE}} = V_{\mathrm{RTE}}(t_2)$, $\ldots$, $V_{\mathrm{RTE}} = V_{\mathrm{RTE}}(t_K)$ from the left, and $U_1 = U_1^{(1)}, U_1 = U_1^{(2)}, \ldots, U_1 = U_1^{(K)}$ from the top, where $U_{1}^{(k)}$ is equal to $U_{1}$ in the $k$th step. The same comments apply to $U_2$.
}
 \label{fig:qc_pite_qaa}
\end{figure*}

\paragraph*{Quantum Amplitude Amplification.}---
Quantum acceleration of ground state preparation may also be realized using probabilistic algorithms by introducing a multi-step scheme and combining it with QAA \cite{Brassard1997, Brassard2000arXiv, Yoder2014PRL, Martyn2021PRXQuantum}.
We consider a $K$-step probabilistic quantum circuit with $K$ ancilla qubits. 
Each ancilla qubit corresponds to a step of the probabilistic algorithm, and all ancilla qubits are measured after the final step.
The output state immediately preceding the measurement is of the form:
\begin{gather}
    |\widetilde{\Psi}\rangle
    =
    a |\Psi_{\mathrm{good}}\rangle
    +
    \sqrt{1-a} |\Psi_{\mathrm{bad}}\rangle
    ,
\label{eq:output_without_QAA}
\end{gather}
where
\begin{gather}
    |\Psi_{\mathrm{good}}\rangle
    =
    \frac{1}{\sqrt{P_{K}}} F_{k}(\mathcal{H})
    |\psi\rangle \otimes |0\rangle^{\otimes K}
\end{gather}
is the desired state with weight $a$
and $|\Psi_{\mathrm{bad}}\rangle$ denotes the orthogonal state of $|\Psi_{\mathrm{good}}\rangle$; 
QAA enhances the coefficients of the $|\Psi_{\mathrm{good}}\rangle$ state by the $m$-times action $\prod_{i=1}^{m} Q(\phi_{2i-1}, \phi_{2i})$ of the following amplitude amplification operator
\begin{gather}
    Q(\phi_{2i-1}, \phi_{2i})
    \nonumber \\
    \equiv
    - \mathcal{U}_{\mathrm{REF}}^{(K)}
    S_{0}^{(n+K)}(\phi_{2i-1})
    \left (\mathcal{U}_{\mathrm{REF}}^{(K)} \right)^{\dagger}
    S_{\chi}(\phi_{2i}),
\label{eq:amplitude_amplification_opr}
\end{gather}
where $S_{\chi}$ denotes an oracle, $S_{0}$ denotes a zero reflection, and $\mathcal{U}_{\mathrm{REF}}^{(K)}$ is defined as $\mathcal{U}_{\mathrm{REF}}^{(K)}|0\rangle^{\otimes (n+K)} = |\widetilde{\Psi}\rangle$.
The rotation angles $\{\phi_i\}$ are chosen as $\phi_i = \pm \pi$ following the conventional method \cite{Brassard2000arXiv, Brassard1997}; however, recently, approaches have been proposed to determine $\{\phi_i\}$ such that the increase in success probability is an approximated sign function \cite{Yoder2014PRL, Martyn2021PRXQuantum}.
The zero reflection and oracle rotate the output state $|\widetilde{\Psi}\rangle$ in the effective two-dimensional space spanned by $|\Psi_{\mathrm{good}}\rangle$ and $|\Psi_{\mathrm{bad}}\rangle$, which are represented by $S_{0}^{(n)}(\phi) = e^{i\phi |0\rangle\langle 0|^{\otimes n}}$ and $S_{\chi}(\phi) = I_{2^n} \otimes S_{0}^{(K)}(\phi) $, respectively.
The circuit depth of the zero-reflection scales at the order of $\mathcal{O}(n)$ with a single ancilla qubit \cite{AdrianoPRA1995}, and Maslov's gate reduces the scaling pre-constant \cite{MaslovPRA2016}.
Figure \ref{fig:qc_pite_qaa} illustrates the amplitude amplification operator $Q$ for the probabilistic algorithm for state preparation in $K$ steps.

The optimal number of repetitions of QAA is derived as follows:
\begin{gather}
    m^{*} = 
    \left\lfloor \frac{(2n+1)\pi}{4 \sin^{-1} a} \right\rfloor.
\end{gather}
Thus, when the total success probability, $P_K$, is low, we execute a first-order Taylor expansion for $\sin^{-1}a$ and obtain the order of optimal repetitions, $m^{*}$, as $m^{*} = \mathcal{O} (1/|c_1|)$.
Accordingly, the computational cost of PITE combined with QAA (henceforth referred to as multi-step PITE) is estimated as follows:
\begin{gather}
    d_{\mathrm{CRTE}} K m^{*}
    =
    \mathcal{O} \left(
        \frac{d_{\mathrm{CRTE}}}{|c_{1}|}
        \ln \left(
            \frac{(1-\delta_K)(1-|c_1|^{2})}{\delta_K |c_1|^2}
        \right) 
    \right),
\end{gather}
where the QAA technique achieves a quadratic acceleration by Eq. (\ref{eq:computational_cost_pite}). 
We also discuss the combined technique comprising PITE and QAA discussed in \cite{Nishi2022arXiv}, where a short-depth circuit is proposed for the first-step PITE and circuit construction is designed for $K$ steps using an ancilla qubit. The multistep PITE method proposed in this study is a natural extension of that in Ref. ~\cite{Nishi2022arXiv} and clearly achieves quadratic acceleration for the whole PITE process.

\paragraph*{Quantum phase estimation}---
QPE is a standard building block, which not only estimates the ground state energy but also prepares the ground state. 
Before comparing multi-step PITE with QPE in terms of numerical results, we briefly discuss QPE.
QPE based on QFT achieves Heisenberg scaling.
Although it typically requires many ancilla qubits, reusing ancilla qubits after measurement and executing subsequent operations depending on the observations enables the same estimation as that using a single ancilla qubit \cite{Griffiths1996PRL, Higgins2007Nature, Berry2009PRA, Ding2023PRXQ}.
For simplicity, standard QPE based on QFT is considered here. 
After each execution of QPE, one of the eigen-energies is loaded onto the ancilla qubits in a binary representation.
We assume the input state for the QPE is $|\psi\rangle = \sum_{i=1}^{N} c_{i}|\lambda_i \rangle$ and $K$ ancilla qubits are used. Using QPE, all eigenvalues \{$\lambda_{i}$\}, which are assumed to be in ascending order, are expressed by a binary representation \{$k_i$\}. In a realistic situation with a finite number of available ancilla qubits, different eigenvalues \{$\lambda_{i}$\} that are energetically close to each other and within an energy resolution $1/T$ can be mapped to the same binary representation $k_i$, where $T\equiv2^{K}$. When we observe $k$ as a binary representation of the estimated eigenvalue, the input state collapses to
\begin{gather}
    |\Psi_{\mathrm{QFT}}\rangle
    =
    \frac{1}{\sqrt{P_{k}}}
    \sum_{i=1}^{N} c_i 
    \alpha_{k|i} |\lambda_{i}\rangle
    ,
\end{gather}
where the state is normalized with probability 
$
    P_{k}
    =
    \sum_{i=1}^{N} |c_{i}|^2 |\alpha_{k|i}|^{2}
$ and periodic function $\alpha_{k|i}$ is defined as 
$
    \alpha_{k|i} 
    \equiv 
    (1/T) \sum_{\tau=0}^{T-1} e^{2\pi i \tau (\lambda_i t_0 -  k )/T},
$
.
Here, $t_0$ is a scaling parameter used to increase or decrease the eigenvalues for precise measurement.
By $T\equiv2^{K}$, we conclude that 
the number of ancilla qubits used is directly proportional to the fineness of the resolution of the eigenvalues. Thus, we take $t_0 = 2^{K-N_C}$, where $N_C = \lfloor \log_{2}(\lambda_{N}-\lambda_{1}) \rfloor$.
Because the number of queries to CRTE increases exponentially, the computational cost for QPE to obtain the ground-state eigenvalue is given by
$
    \mathcal{O}\left(
        1/ (\sqrt{\delta_K} |c_1|^{2}),
    \right) 
$
~ \cite{Ge2019JMP, SM}.
By comparing it with Eq. (\ref{eq:computational_cost_pite}), we emphasize that PITE improves computational cost exponentially with respect to the infidelity $\delta_K$ over QPE.

\paragraph*{Numerical results.}---
The numerical simulation demonstrates the strengths of the proposed method, which is implemented using Qiskit, an open-source library for quantum simulations \cite{Qiskit}. 
The Heisenberg model is adopted as the computational model.
\begin{gather}
    \mathcal{H}
    =
    \sum_{\langle j,k \rangle} \vec{\sigma}_{j} \cdot \vec{\sigma}_{k}
    +
    \sum_{j} h_j \sigma_{j}^{z}
\end{gather}
where $\vec{\sigma}_{j} = (\sigma_{j}^{x}, \sigma_{j}^{y}, \sigma_{j}^{z})$ is the Pauli matrix acting on the $j$th spin and $\langle j, k \rangle$ represents the combination of the nearest neighbors of the closed one-dimensional chain. 
$h_j$ represents the strength of the magnetic field, which is randomly selected from a uniform distribution $h_j \in [-1, 1]$.
The CRTE gate is implemented using a fourth-order Trotter decomposition \cite{Suzuki1991JMP} for even-odd groups of the Hamiltonian \cite{Childs2019PRL, Childs2021PRX}. 
The dependence of the accuracy and total computational complexity on the order of the Trotter decomposition is numerically demonstrated in SM \cite{SM}.
The muti-step PITE method is adopted as the probabilistic algorithm while preparing the ground state. 
The PITE method employs a constant energy shift to increase the total success probability, and adopts computationally efficient scheduling of the imaginary-time step size \cite{Nishi2023arXiv}.

\begin{figure}
        \centering
        \includegraphics[width=0.45 \textwidth]{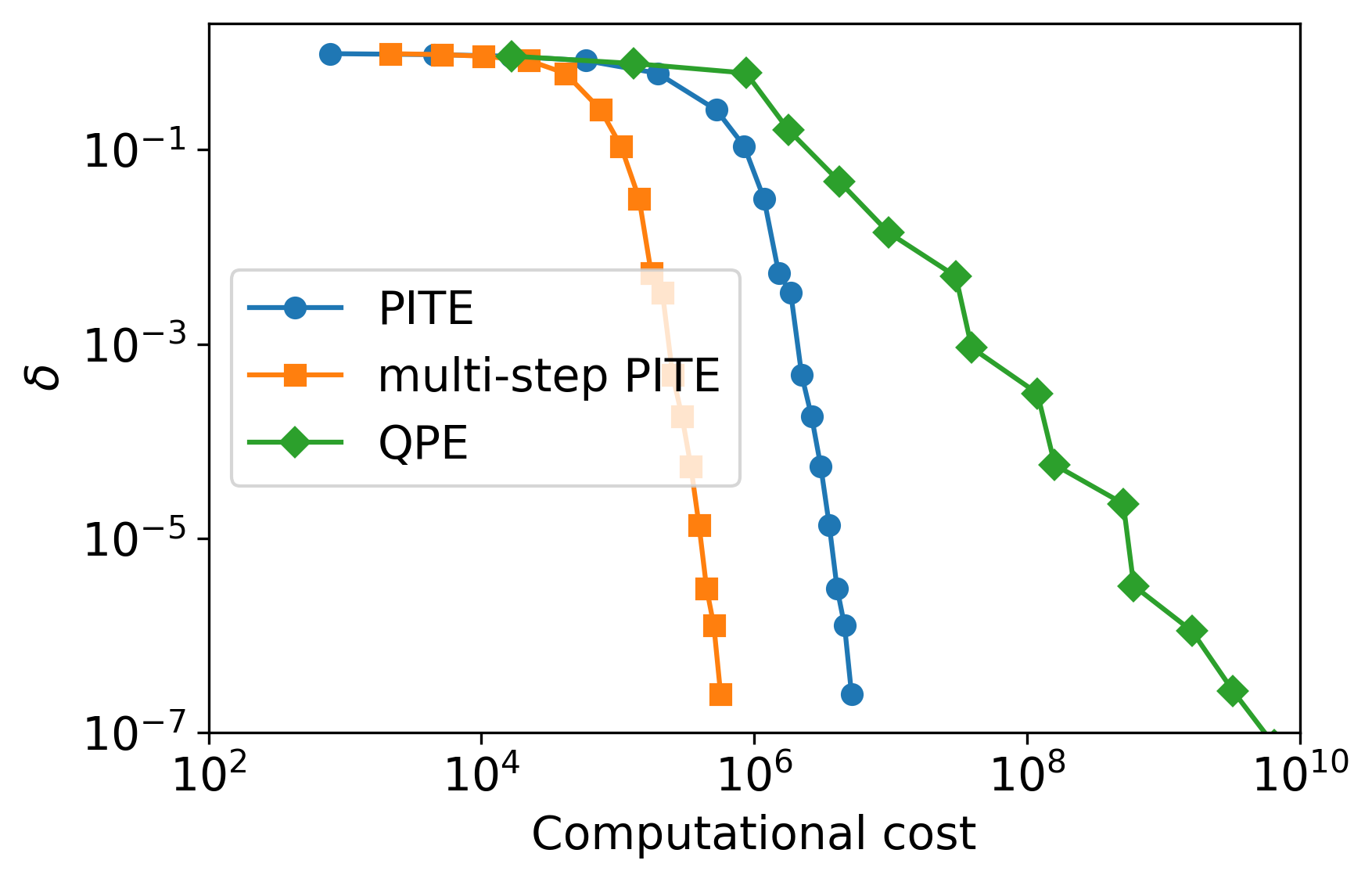}
\caption{
Plots of the infidelity $\delta_K$ as a function of computational cost for PITE, multi-step PITE, and QPE, respectively. 
The results are obtained using the one-dimensional Heisenberg chain with eight spins.
}
 \label{fig:rslt_complexity}
\end{figure}

Figure \ref{fig:rslt_complexity} illustrates the infidelity $\delta_K$ as a function of the computational costs for QPE, PITE, and multi-step PITE.
The initial state is selected as a uniform probability weight with respect to each eigenvector; i.e., $|c_i|^{2} = 1/N$. 
As the number of steps in PITE or the number of digits, i.e., the resolution in QPE increases, the infidelity decreases, whereas the computational cost increases.
As the infidelity decreases, the success probability decreases and approaches $1/N$. This behavior is common in PITE and QPE.
PITE and QPE differ in the number of queries in the CRTE block.
In QPE, the number of queries increases exponentially with the number of ancilla qubits; thus, the computational cost increases linearly with infidelity.
However, in PITE, the number of queries increases linearly with the step, implying that the scaling of the computational cost is logarithmic with respect to infidelity.
The figure confirms that PITE is exponentially faster than QPE with respect to infidelity $\delta_K$, as given by Eq. ~ (\ref{eq:computational_cost_pite}).
Further, in the case of multi-step PITE, we observe an increase in the computational cost due to the quadratic acceleration with respect to the success probability, despite the overhead incurred due to zero reflections.
We verify the method with different orders and use a fourth-order trotter, which yields results in good agreement with the exact solution \cite{SM}.

\begin{figure}
        \centering
        \includegraphics[width=0.45 \textwidth]{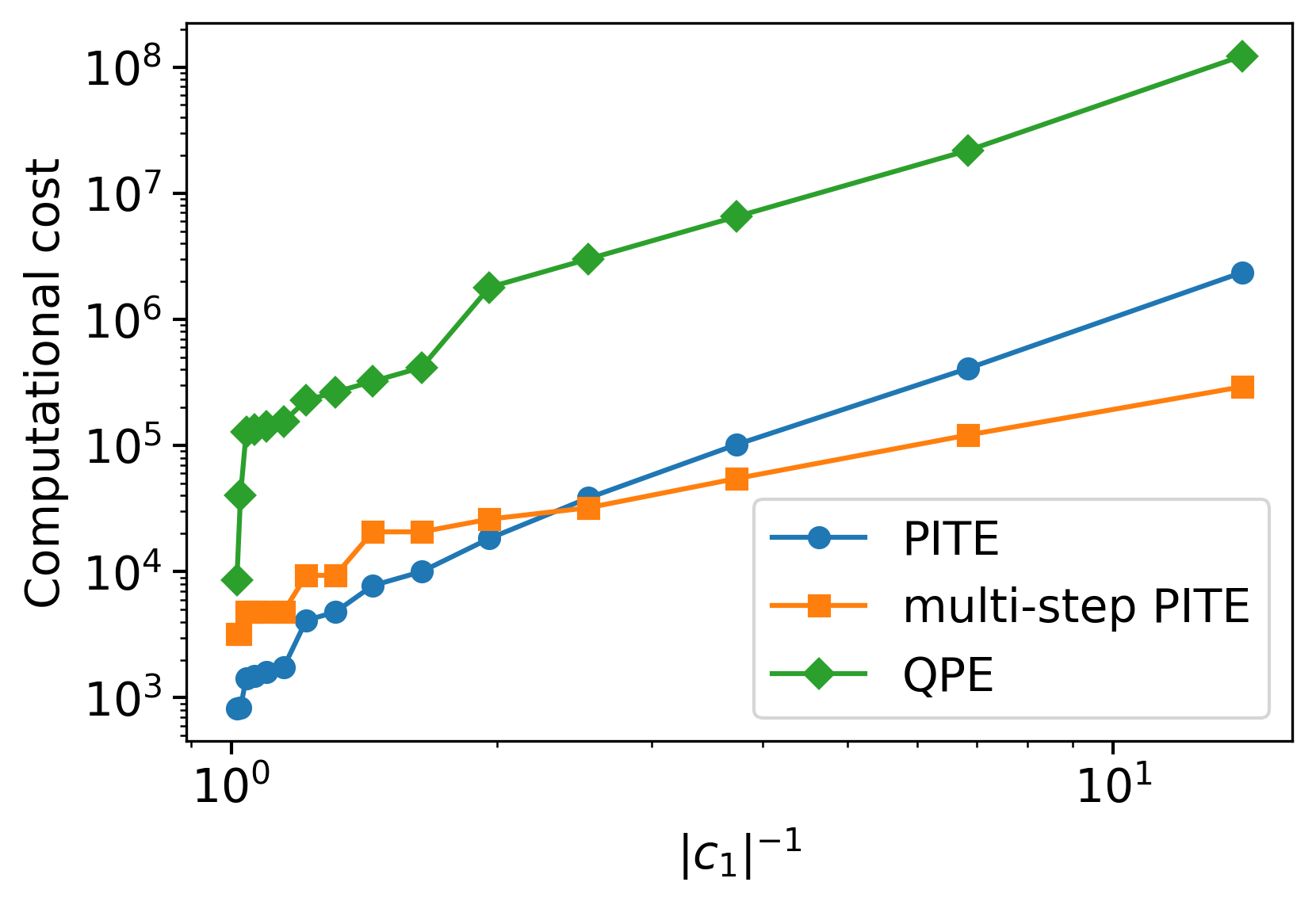}
 \caption{
Plots of computational cost for PITE, multi-step PITE, and QPE as functions of the inverse probability weight of the ground state in the input state.
}
 \label{fig:rslt_complexity_weight}
\end{figure}

To demonstrate quadratic acceleration with respect to the probability weight $|c_1|^2$ of the ground state clearly in the case of multi-step PITE, the dependence of computational cost on $|c_1|^2$ is plotted in Figure \ref{fig:rslt_complexity_weight}.
The computational cost is estimated when the infidelity is below $\delta_K = 10^{-4}$.
The probability weight of the initial state is determined using a Gaussian distribution as a function of the eigenvalues corresponding to the mean ground-state energy.
The probability weights of the ground states are increased by changing the variance of the Gaussian distribution.
The computational cost of PITE is observed to be lower than that of QPE over the entire region, owing to the exponential advantage of PITE in terms of the infidelity, $\delta_K$. 
Although the computational cost of multi-step PITE is higher than that of PITE without QAA at $|c_1| \approx 1/2$ because of the overhead induced by QAA, it decreases as $|c_1|^{-1}$ increases when $|c_1| < 1/2$. 
This confirms quadratic acceleration induced by QAA. 
The crossing point of the computational costs of PITE and multi-step PITE is determined by the relationship between the circuit depths of PITE and zero reflection \cite{SM}.
As the computational overhead induced by zero reflection decreases, QAA increases the computational cost over a wide range.

\paragraph*{Conclusions.}---
In this letter, we propose a quantum algorithm for ground state preparation that offers several quantum advantages. 
The recently proposed PITE method, comprising a single ancilla qubit and forward and backward CRTE operations, calculates the ground state non-variationally on a quantum computer. 
Although PITE exhibits an exponential advantage over QPE in terms of the infidelity $\delta$, its probabilistic nature degrades the computational cost by a scale of order $\mathcal{O}(|c_1|^{-2} \log \delta^{-1}),$ where $|c_1|^2$ denotes the probability weight of the ground state in the initial state.
Here, we combine QAA with PITE, achieving quadratic acceleration compared to the classical method, where delayed measurement enables multi-step amplitude amplification. 
Numerical simulations emphasize the strengths of the proposed algorithm, which is implemented on a fault-tolerant quantum computer (FTQC). 
However, the development of quantum algorithms for early FTQC and post-noisy intermediate-scale quantum devices remains an important research topic.
In this context, reducing the number of ancilla qubits and the circuit depth of the proposed algorithm is a promising direction for future research.  
In addition, preparing a good initial state using classical pre-processing techniques is expected to contribute to fast quantum-state preparation irrespective of the application of QAA. 
By combining the proposed algorithm, various physical quantities such as the one-body Green's function \cite{kosugi2020PRA}, the linear response function \cite{kosugi2020PRR}, and microcanonical and canonical properties \cite{Lu2021PRXQ} can be calculated.
This method is also applicable to various problems such as the optimization of the structure geometry based on an exhaustive search \cite{Kosugi202210arXiv} and investigating an electron under a magnetic field \cite{Kosugi2023JJAP}. 
Thus, this study contributes to the gamut of numerical simulations related to material sciences that are executable on quantum computers.

This work was supported by MEXT under "Program for Promoting Researches on the Supercomputer Fugaku" (JPMXP1020200205) and by JSPS KAKENHI under Grant-in-Aid for Scientific Research (A) No. 21H04553.

\bibliographystyle{apsrev4-2}
\bibliography{ref}

\end{document}


\pagebreak
\widetext
\begin{center}
\textbf{\large Supplementary Material: Quadratic accelration of multi-step probabilistic algorithms for state preparation}
\end{center}
\setcounter{equation}{0}
\setcounter{figure}{0}
\setcounter{table}{0}
\setcounter{page}{1}
\makeatletter
\renewcommand{\theequation}{S\arabic{equation}}
\renewcommand{\thefigure}{S\arabic{figure}}
\renewcommand{\thesection}{S\arabic{section}}
\renewcommand{\bibnumfmt}[1]{[S#1]}
\renewcommand{\citenumfont}[1]{S#1}

\section{Implementation of non-unitary operator in a probabilistic manner}
\subsection{Probabilistic imaginary-time evolution method}
\subsubsection{Exact non-unitary operator}
In our previous work \cite{Kosugi2022PRR, Nishi2022arXiv}, we presented a quantum circuit for the probabilistic operation of the non-unitary Hermitian operator $f(\mathcal{H})$, where $\mathcal{H}$ denotes the Hamiltonian of an $n$-qubit system.
The quantum circuits for the nonunitary operation $f(\mathcal{H})$ are depicted in Fig. \ref{circuit:probabilistic_algorithms_state_preparation}(a), comprising a single ancilla qubit and two controlled unitary gates, $e^{\pm \kappa \Theta}$. 
The generators of the unitary gates are denoted by $\pm \kappa \Theta$ where
\begin{gather}
    \Theta 
    \equiv
    \arccos \left[
        \frac
            {
                f (\mathcal{H})
                +
                \sqrt{1-\left[f (\mathcal{H})\right]^{2}}
            }
            {\sqrt{2}} 
    \right] ,
\label{eq_sm:exact_ite_Theta}
\end{gather}
is a Hermitian operator for an $n$-qubit system and
\begin{gather}
    \kappa
    \equiv
    \operatorname{sgn} \left(
        \|f(\mathcal{H})\|
        -
        \frac{1}{\sqrt{2}}
    \right) ,
\end{gather}
is defined, yielding:
$
    \cos \Theta
    =
    (f(\mathcal{H}) + \sqrt{1 - f^{2}(\mathcal{H})})
$
and 
$
    \sin \kappa \Theta
    =
    (f(\mathcal{H}) - \sqrt{1 - f^{2}(\mathcal{H})})
$.
The single-qubit gate shown in Fig. \ref{circuit:probabilistic_algorithms_state_preparation}(a) is defined as follows:
\begin{gather}
    W 
    \equiv
    \frac{1}{\sqrt{2}}
    \begin{pmatrix}
        1 & -i \\
        1 & i
    \end{pmatrix} .
\end{gather}

An interesting nonunitary operator is the imaginary time-evolution (ITE) operator, defined as follows:
\begin{gather}
    f(\mathcal{H})
    =
    \Gamma e^{-\tau \mathcal{H}} ,
\label{eq:f_exact_pite}
\end{gather}
where $\tau$ denotes imaginary time and $\Gamma$ is an adjustable parameter.
The ITE operator acts successfully on the target qubits when an ancilla qubit is observed in the $|0\rangle$ state. 
In contrast, when an ancilla qubit is observed as $|1\rangle$ state, the target qubits occupy undesired states. 
Because the state acted upon by the ITE operator is obtained probabilistically, this method is called the probabilistic imaginary-time evolution (PITE) method.

\subsubsection{Approximate non-unitary operator}
\label{sec_sm:approximate_nonunitary}
The implementation method for unitary $e^{\pm i\kappa \Theta}$ is not straightforward; thus, we approximate it using the first order of $\mathcal{H}$. 
To this end, we first decompose the non-unitary $f(\mathcal{H})$ into a product consisting of $K$ factors, such that the approximation is well-constructed.
$f(\mathcal{H})$ is rewritten as $F_K(\mathcal{H})$, which explicitly indicates the parameter $K$ and each factor is denoted by $f_k(\mathcal{H})$: 
$
    F_K(\mathcal{H})
    =
    \prod_{k=1}^{K} f_k(\mathcal{H})
$.
To distinguish between variables at different steps, we denote the variable at the $k$th step using the subscript $k$.
Executing the first-order Taylor decomposition yields:
\begin{gather}
    \kappa \Theta_k
    =
    a^{(k)}_0 + a^{(k)}_1 \mathcal{H} 
    +
    \mathcal{O}\left(\mathcal{H}^{2}\right),
    \label{PITEwithQAA:Theta_taylor_1st_order}
\end{gather}
where $a^{(k)}_0$ and $a^{(k)}_1$ denote the coefficients of the Taylor expansion, which depend on the type of non-unitary $f_k(\mathcal{H})$.

Let us consider the ITE operator $\Gamma e^{-\tau \mathcal{H}}$ as an example. First, the imaginary time $\tau$ is divided into small fragments.
$
    \tau 
    \equiv
    \sum_{k=1}^{K} \Delta \tau_{k}    
$.
The ITE operator for imaginary time $\tau$ is decomposed as follows:
\begin{gather}
    \Gamma e^{-\tau \mathcal{H}}
    =
    \prod_{k=1}^{K} \gamma_{k} e^{-\Delta\tau_{k} \mathcal{H}} ,
\end{gather}
where $\Gamma = \prod_{k=1}^{K} \gamma_{k}$, 
$\gamma_k$ is an adjustable parameter that satisfies the conditions $0 < \gamma < 1$ and $\gamma \neq 1/\sqrt{2}$ and is introduced to avoid singularities.
The ITE operator for a small imaginary time $\gamma_{k} e^{-\Delta\tau_{k} \mathcal{H}}$ is implemented as described above. 
When the arccosine function in Eq. (\ref{eq_sm:exact_ite_Theta}) is intended for $f_k(\mathcal{H}) = \gamma_{k} e^{-\Delta\tau_{k} \mathcal{H}}$, the coefficients of the Taylor expansion are calculated as follows:
$
    a_0^{(k)}
    \equiv 
    \kappa \arccos \left[ (\gamma_{k}+\sqrt{1-\gamma_{k}^{2}})/\sqrt{2} \right]
$
and
$
    a_1^{(k)} \equiv \gamma_{k}/\sqrt{1-\gamma_{k}^{2}}
$. 
Consequently, the approximated non-unitary operator in the first order of $\Delta\tau_k$ is given by
\begin{gather}
    f_{k}(\mathcal{H})
    =
    \gamma_{k} \left[
        \cos (\mathcal{H} \Delta\tau_{k} s_{k})
        - 
        \frac{1}{s_{k}} \sin (\mathcal{H} \Delta\tau_{k} s_{k} )
    \right]
    \nonumber \\
    =
    \sin (-\mathcal{H} \Delta\tau_{k} s_{k} + \varphi_{k} )
    ,
\end{gather}
where $s_{k} = \tan \varphi_{k}$. 

A quantum circuit of the approximate PITE operator is depicted in Fig. \ref{circuit:probabilistic_algorithms_state_preparation}(b). 
It comprises forward and backward controlled real-time evolution (CRTE) operators. 
CRTE gates can be implemented efficiently; in particular, sophisticated implementation methods can be utilized in the context of Hamiltonian simulations.

\subsubsection{Computational cost}
We now consider the computational cost of preparing the ground state of the Hamiltonian $\mathcal{H}$ using multiple steps of the approximate nonunitary operator described in Section \ref{sec_sm:approximate_nonunitary}.
$K$ ancilla qubits are prepared, each of which is used in the PITE step.
The input state, including ancilla qubits, is given by $|\psi\rangle \otimes |0\rangle^{\otimes K}$. 
The action of $K$ steps of the quantum circuit corresponding to the approximate non-unitary operator yields:
\begin{gather}
    F_K(\mathcal{H}) |\psi\rangle
    \otimes 
    |0\rangle^{\otimes K}
    +
    \mathrm{(other ~ states)}.
\end{gather}

During the measurement of all ancilla qubits in the $|0\rangle^{\otimes K}$ state, the entangled wave function collapses to 
\begin{gather}
    |\Psi_{K} \rangle
    =
    \frac{1}{\sqrt{P_{K}}} F_{K}(\mathcal{H}) |\psi\rangle ,
\end{gather}
where $P_{K}$ denotes the total success probability for all PITE steps.
When the initial target-qubits state is expanded as
\begin{gather}
    |\psi\rangle
    =
    \sum_{i=1}^{N} c_i |\lambda_{i} \rangle,
\end{gather}
the infidelity, $\delta_K$, for $K$ steps is calculated as follows:
\begin{gather}
    \delta_K 
    =
    1 - |\langle \lambda_{1} | \Psi_{K}\rangle|^{2}
    =
    1 - \frac{1}{P_K}|c_1|^{2} F_{K}^{2}(\lambda_1).
\end{gather}
Based on the aforementioned relation, the total success probability can be expressed as follows:
\begin{gather}
    P_K = \frac{1}{1-\delta_K} |c_1|^{2} F_K^{2}(\lambda_1) .
\label{eq_sm:total_success_prob}
\end{gather}
$F_K^{2}(\lambda_1)$ denotes the $k$-product of $f_k^{2}(\lambda_1)$. 
$
    F_K^{2}(\lambda_1) 
    = 
    \prod_{k=1}^{K} f_k^{2}(\lambda_1)
$ 
and $f_k^{2}(\lambda_1)$ assumes real values in the range, $[0, 1]$. 
Thus, the total success probability decays exponentially with the progression of PITE steps.
This exponential decay can be avoided by shifting the origin of the Hamiltonian in a way such that $F_K^{2}(\lambda_1) = 1$. 
In approximate PITE, this constant energy shift can be achieved by
\begin{gather}
    E_{k}
    =
    \lambda_{1}
    -
    \frac{1}{\Delta\tau_{k}s_{k}} \left[
        \tan^{-1} s_{k} 
        - 
        \frac{\pi}{2}(2n+1)
    \right] ,
\label{eq:constant_energy_shift}
\end{gather}
with an integer $n$ for $\lambda_{i} \to \lambda_{i} - E_{k}$.
Correspondingly, in the quantum circuit of approximate PITE [Fig. \ref{circuit:probabilistic_algorithms_state_preparation}(b)], we change the rotation angle of the $R_{z}$ gate using the relation $2\theta_{k} \to 2\theta_{k} + 2E_{k}s_{k}\Delta\tau_{k}$.

The total success probability can be expressed in a different form compared to Eq. (\ref{eq_sm:total_success_prob}) as follows:
\begin{gather}
    P_K 
    = 
    |c_1|^{2} F_K^{2}(\lambda_1) 
    +
    \sum_{i>1} |c_i|^{2} F_K^{2}(\lambda_i) .
\label{eq_sm:total_success_prob_another}  
\end{gather}
From Eqs. (\ref{eq_sm:total_success_prob}) and (\ref{eq_sm:total_success_prob_another}), respectively,
we have
\begin{gather}
    \frac{\delta_K}{1-\delta_K}
    =
    \sum_{i>1}  \frac{|c_i|^{2} F_K^{2}(\lambda_i) }{ |c_1|^{2} F_K^{2}(\lambda_1) }.
\label{eq_sm:error_to_be_evaluated}
\end{gather}
The right-hand side of Eq. (\ref{eq_sm:error_to_be_evaluated}) was evaluated in detail in Ref. \cite{Nishi2023arXiv}, where the left-hand side of Eq. ~ (\ref{eq_sm:error_to_be_evaluated}) was expressed as $\widetilde{\varepsilon} = \delta_K / (1 - \delta_K)$.
$\widetilde{\varepsilon}$ denotes an error, defined as follows: 
$
    \widetilde{\varepsilon}
    \equiv
    \varepsilon (4-\varepsilon)
    /
    (2-\varepsilon)^{2} 
$ with
$
    \varepsilon
    \equiv
    \left \|
        \Gamma e^{-\mathcal{H}}\tau|\psi\rangle
        -
        F_K(\mathcal{H})|\psi\rangle,
    \right \|^2
$.
In conclusion, if $\Delta \tau_k$ is selected to vary linearly with the step and a constant energy shift is adopted for Eq. (\ref{eq:constant_energy_shift}), the number of steps required to achieve infidelity $\delta_K$ can be estimated as follows: 
\begin{gather}
    K 
    =
    \mathcal{O} \left(
    \ln \left(
        \frac{(1-\delta_K)(1-|c_1|^{2})}{\delta_K |c_1|^2}
    \right) \right) .
\label{eq:steps_linear_scaling}
\end{gather}
Finally, we arrive at the following equation for the computational cost of PITE as presented in the main text:
\begin{gather}
    \frac{d_{\mathrm{CRTE}} K}{P_{K}} 
    =
    \mathcal{O} \left(
        \frac{d_{\mathrm{CRTE}}}{|c_{1}|^{2}}
        \ln \left(
            \frac{(1-\delta_K)(1-|c_1|^{2})}{\delta_K |c_1|^2}
        \right) 
    \right) ,
\label{eq_sm:computational_cost_pite}
\end{gather}
where $d_{\mathrm{CRTE}}$ denotes the circuit depth of the CRTE operator.

\subsection{Cosine propagation}
\begin{figure}
    \centering
    \includegraphics[width=0.9 \textwidth]{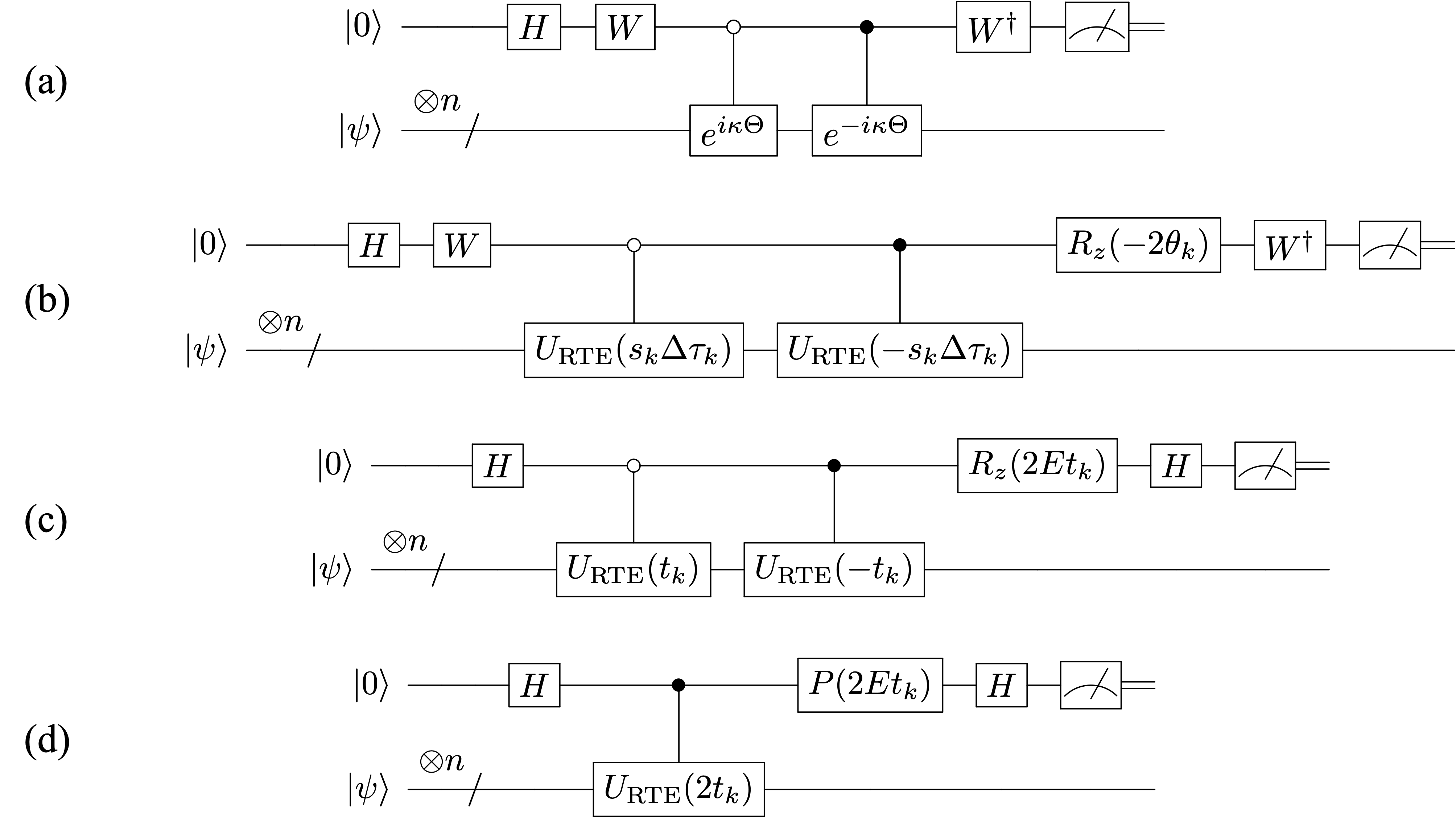} 
\caption{
(a) Quantum circuit, $\mathcal{C}_{\mathcal{M}}$, for the probabilistic operation of a non-unitary $\mathcal{M}$ that acts on an input $n$-qubit state, $|\psi\rangle$. 
(b) Quantum circuit, $\mathcal{C}_{\mathrm{PITE}}^{(1)}$, is equivalent to $\mathcal{C}_{\mathrm{PITE}}$ in the first order of $\Delta \tau$.
(c) Quantum circuit, $\mathcal{C}_{\mathrm{cos}}$, for cosine propagation. 
(d) Quantum circuit equivalent to $\mathcal{C}_{\mathrm{cos}}$ realized using a single CRTE gate while ignoring the global phase. 
$H$ and $Z({\theta})$ represent Hadamard and phase gates, respectively.
}
\label{circuit:probabilistic_algorithms_state_preparation}
\end{figure}

Previous studies \cite{Ge2019JMP, Choi2021PRL, Lu2021PRXQ, Meister2022arXiv} established quantum circuits for the non-unitary cosine function of the Hamiltonian $\mathcal{H}$ as follows: 
\begin{gather}
    f^{(\mathrm{cos})}_{k} (\mathcal{H})
    =
    \cos \left[
        (\mathcal{H} - E) t_{k}
    \right],
\end{gather}
This is depicted in Fig. \ref{circuit:probabilistic_algorithms_state_preparation}(c), where 
$Z(\theta) = Z_{\theta} = |0\rangle\langle 0| + e^{i\theta} |1\rangle\langle 1|$ denotes the phase gate.
Multiple operations using the cosine function decrease the weights of states whose eigenvalues are not equal to a given energy, $E$. 
Ref. \cite{Choi2021PRL} used only a single CRTE gate, as depicted in Fig. \ref{circuit:probabilistic_algorithms_state_preparation}(d), which also provided the cosine propagation ignoring the global phase, as follows:
\begin{gather}
    \tilde{f}^{(\mathrm{cos})}_{k} (\mathcal{H})
    =
    \frac{1}{2} \left[
        I_{2^n} + e^{-2i (\mathcal{H}-E) t_k}
    \right]
    =
    e^{-i (\mathcal{H} - E) t_{k}}
    \cos \left[
        (\mathcal{H} - E) t_{k}
    \right] .
\end{gather}

\section{Quantum phase estimation based on quantum Fourier transformation}
In this section, we review the gate complexity for the preparation of the ground state using QPE for known ground energies, as presented in Proposition 3 in Ref. \cite{Ge2019JMP}. 
The quantum circuit for a QPE based on the quantum Fourier transformation (QFT) \cite{Nielsen2000Book} changes the initial target qubits state as follows:
$
    |\psi\rangle 
    = 
    \sum_{i=1}^{N} c_i |\lambda_{i}\rangle
$
and the $K$ ancillae qubits to
\begin{gather}
    |\psi\rangle \otimes |0\rangle^{\otimes K}
    \to
    \sum_{k=0}^{T-1}
    \underbrace{
        \sum_{i=1}^{N} c_i 
        \alpha_{ik} |\lambda_{i}\rangle  
    }_{\equiv |\Phi_{k} \rangle}
    \otimes |k\rangle ,
\end{gather}
where 
\begin{gather}
    \alpha_{ik}
    \equiv
    \frac{1}{T} \sum_{\tau=0}^{T-1}
    e^{2\pi i \tau (\lambda_i t_0 - k) /T}
    =
    \frac{1}{T} 
    \left(
        \frac{1 - e^{2\pi i(T\lambda_i t_0 - k)}}{1 - e^{2\pi i(\lambda_i t_0 - k/T)}}
    \right) ,
\end{gather}
and $T \equiv 2^K$.
$t_0$ denotes a scaling parameter that enlarges or reduces the search range of eigenvalues introduced for precise estimation.
The post-selection state with respect to the eigenvalue $k$ in the binary representation is given by
\begin{gather}
    |\Psi_{\mathrm{QFT}}\rangle
    =
    \frac{1}{\sqrt{P_{k}}}
    \sum_{i=1}^{N} c_i 
    \alpha_{i k} |\lambda_{i}\rangle ,
\end{gather}
where the state is normalized with respect to the success probability, expressed as follows:
\begin{gather}
    P_{k} 
    =
    \sum_{i=1}^{N} |c_i|^{2} |\alpha_{ik}|^{2} .
\end{gather}
The infidelity of $|\Psi_{\mathrm{QFT}}\rangle$ is calculated as follows:
\begin{gather}
    \delta
    =
    1 - |\langle \lambda_{1} | \Psi_{\mathrm{QFT}}\rangle|^{2}
    =
    \frac
        {\sum_{i > 1} |c_i|^2 |\alpha_{ik}|^2 }
        {\sum_{i=1}^{N} |c_i|^2 |\alpha_{ik}|^2} .
\label{eq_sm:infidelity_qpe}
\end{gather}
The success probability is rewritten in terms of infidelity as
\begin{gather}
    P_k
    =
    \frac{1}{1- \delta} |c_1|^2 |\alpha_{1k}|^2 .
\label{eq_sm:tot_prob_with_infidelity}
\end{gather}

To estimate the order of the scaling of $K$ with respect to $\delta$, we investigate the inequality for $\alpha_{ik}$. 
Using $|1 - e^{i\theta}| \geq 2$ for $\theta \in \mathbb{R}$, we obtain
\begin{gather}
    |\alpha_{ik}|
    \leq
    \frac{2}{T |1 - e^{2\pi i(\lambda_i t_0 - k/T)}|} .
\end{gather}
For $\theta \in [-\pi, \pi]$, we know that
\begin{gather}
    \frac{2}{\pi}|\theta| 
    \leq
    |1 - e^{i\theta}|
    \leq 
    |\theta| ,
\end{gather}
which yields the following relationship \cite{Nielsen2000Book}:
\begin{gather}
    \frac{1}{\pi T \Delta}
    \leq
    |\alpha_{ik}|
    \leq
    \frac{1}{2T \Delta}, 
\label{eq_sm:qpe_inequality_alpha_ik}
\end{gather}
where $\Delta \equiv |\lambda_i t_0 - k/T|$.
In addition, when $k$ satisfies $\Delta = |\lambda_i t_0 - k/T| < 1/2T$, \cite{Ge2019JMP}
\begin{gather}
    \frac{2}{\pi}
    \leq
    |\alpha_{ik}| .
\label{eq_sm:qpe_inequality_alpha_0k}
\end{gather}
Using the two inequalities (\ref{eq_sm:qpe_inequality_alpha_ik}) and (\ref{eq_sm:qpe_inequality_alpha_0k}), the infidelity, $\delta$, in Eq. (\ref{eq_sm:infidelity_qpe}) is demonstrated to be bounded, as follows:
\begin{gather}
    \delta
    \leq 
    \frac
        {\sum_{i > 1} |c_i|^2 |\alpha_{ik}|^2 }
        {|c_1|^2 |\alpha_{1k}|^2}
    \leq
    \left(
        \frac{\pi}{4 |c_1| T \Delta}
    \right)^{2}
    (1 - |c_1|^2)
    \leq
    \left(
        \frac{\pi}{4 |c_1| T \Delta}
    \right)^{2} .
\end{gather}
To achieve the infidelity, $\delta$, it is sufficient to consider $K$ as follows:
\begin{gather}
    2^{K}
    =
    \mathcal{O}\left(
        \frac{1}{\sqrt{\delta} |c_1| \Delta}
    \right) .
\end{gather}
The number of queries of CRTE scales with an order of $\mathcal{O}(2^K)$. 
Because the controlling unitary gate of QPE is a CRTE gate, a Hamiltonian simulation must be performed in real time and multiplied by $\mathcal{O}(2^K)$. As a simple case, we implement a CRTE gate by dividing it into $\mathcal{O}(2^K)$ parts:
The computational cost of QFT is relatively small compared to that of CRTE sequences; thus, the depth of the quantum circuit for QPE is expressed as follows:
\begin{gather}
    d_{\mathrm{QPE}}
    =
    \mathcal{O}(2^K d_{\mathrm{CRTE}})
    =
    \mathcal{O}\left(
        \frac{d_{\mathrm{CRTE}}}{\sqrt{\delta} |c_1| \Delta}
    \right) .
\end{gather}
Considering the success probabilities in Eq. (\ref{eq_sm:tot_prob_with_infidelity}), the computational cost of QPE is given by
\begin{gather}
    \frac{d_{\mathrm{QPE}}}{P_k}
    =
    \mathcal{O}\left(
        \frac
            {d_{\mathrm{CRTE}}(1-\delta)}
            {\sqrt{\delta} |c_1|^3 \Delta}
    \right) ,
\end{gather}
where we assume $|\alpha_{1k}| = \mathcal{O}(1)$.
Importantly, the computational cost of QPE for the calculation of the ground state energy scales with an order of $\mathcal{O}\left(1/|c_1|^3\right) $.
It should be noted that a suitable amplitude amplification technique reduces the computational cost in \cite{Ge2019JMP}
\begin{gather}
    \frac{d_{\mathrm{QPE}}}{P_k}
    =
    \mathcal{O}\left(
        \frac
            {d_{\mathrm{CRTE}}(1-\delta)}
            {\sqrt{\delta} |c_1|^2 \Delta}
    \right) ,
\end{gather}
which leads to scaling of the order of $\mathcal{O}\left(1/|c_1|^2\right)$.

\section{Quantum circuit for controlled-real-time evolution}
\label{sec_sm:qc_for_rte}
\subsection{Lie-Trotter-Suzuki decomposition}
\label{sec_sm:qc_for_rte_trotter}
The Hamiltonian for the closed one-dimensional Heisenberg model discussed in the main text is given by
\begin{gather}
    \mathcal{H}
    =
    \sum_{\langle j,k \rangle} \vec{\sigma}_{j} \cdot \vec{\sigma}_{k}
    +
    \sum_{j} h_j \sigma_{j}^{z}, 
\end{gather}
where $\vec{\sigma}_{j} = (\sigma_{j}^{x}, \sigma_{j}^{y}, \sigma_{j}^{z})$, and $\langle j, k \rangle$ represent the combination of the nearest neighbors of the closed one-dimensional chain. 
$h_j$ represents the strength of the magnetic field, which is randomly selected from a uniform distribution, $h_j \in [-1, 1]$.
The Hamiltonian is divided into two groups in an even-odd manner \cite{Childs2019PRL, Childs2021PRX}.
\begin{gather}
    \mathcal{H}
    =
    \mathcal{H}_{1} 
    +
    \mathcal{H}_{2} ,
\end{gather}
where
\begin{gather}
    \mathcal{H}_{1} 
    =
    \sum_{j=1}^{\lfloor n/2 \rfloor}
    \left(
        \vec{\sigma}_{2j-1} \cdot \vec{\sigma}_{2j}
        +
        h_{2j-1} \sigma^{z}_{2j-1}
    \right)
    \nonumber \\
    \mathcal{H}_{2} 
    =
    \sum_{j=1}^{\lceil n/2 \rceil - 1}
    \left(
        \vec{\sigma}_{2j} \cdot \vec{\sigma}_{2j+1}
        +
        h_{2j} \sigma^{z}_{2j}
    \right) .
\end{gather}
Although all summands in the sub-Hamiltonian commute with each other, $\mathcal{H}_{1}$ and $\mathcal{H}_{2}$ do not: $[\mathcal{H}_{1}, \mathcal{H}_{2}] \neq 0$.
In other words, the RTE for the sub-Hamiltonian is decomposed exactly as follows: 
\begin{gather}
    e^{i\mathcal{H}_1 t}
    =
    \prod_{j=1}^{\lfloor n/2 \rfloor}
    U_{2j-1, 2j}
    , ~~
    e^{i\mathcal{H}_2 t}
    =
    \prod_{j=1}^{\lceil n/2 \rceil - 1}
    U_{2j, 2j+1} ,
\end{gather}
where $U_{ij}$ denotes the two-qubit gate unitary operation.
\begin{gather}
    U_{ij}
    =
    \exp\left[
        it \left(
            \vec{\sigma}_{2j} \cdot \vec{\sigma}_{2j+1}
            +
            h_{2j} \sigma^{z}_{2j}
        \right)
    \right] .
\end{gather}

The implementation of the RTE for the total Hamiltonian $\mathcal{H} = \mathcal{H}_{1} + \mathcal{H}_{2}$ is realized using the Li–-Trotter-–Suzuki formula \cite{Suzuki1991JMP}:
\begin{gather}
    \mathcal{S}_{1}(t)
    =
    e^{it H_1} e^{it H_2} + \mathcal{O}(t^{2})
    \\
    \mathcal{S}_{2}(t)
    =
    e^{i(t/2) H_1} e^{it H_2} e^{i(t/2) H_1} + \mathcal{O}(t^{3})
    \\
    \mathcal{S}_{2k}(t)
    =
    \mathcal{S}_{2k-2}^{2}(u_{k}t)
    \mathcal{S}_{2k-2}((1-4u_{k})t)
    \mathcal{S}_{2k-2}^{2}(u_{k}t)
    + 
    \mathcal{O}(t^{2k+1}) ,
\end{gather}
where 
\begin{gather}
    u_{k}
    \equiv
    \frac{1}{4-4^{\frac{1}{2k-1}}} .
\end{gather}
The even-order Trotter decomposition, $\mathcal{S}_{2k}(t)$, is derived recursively using the previous even operator, $\mathcal{S}_{2k-2}(t)$. 
The circuit depth for $\mathcal{S}_{2k}(t)$ is given by 
$
    d_{\mathcal{S}_{2k}} = 5^{k-1} d_{\mathcal{S}_{2}},
$.

Consider the simulation of the RTE of an operator $\mathcal{H} = \sum_{\gamma=1}^{\Gamma} \mathcal{H}_\gamma$ comprising an anti-Hermitian operator, $\mathcal{H}_\gamma$. 
For a long-term simulation, the simulation time $t \geq 0$ is divided into $r$ steps, and the $p$th order Trotter--Suzuki decomposition is applied:
The following accuracy is achieved: 
\begin{gather}
    \left\|
        \mathcal{S}_{p}^{r} \left(\frac{t}{r}\right)
        -
        e^{t\mathcal{H}}
    \right\|
    =
    \mathcal{O}(\epsilon) ,
\end{gather}
using the Trotter number, $r$ \cite{Childs2021PRX}, given by 
\begin{gather}
    r 
    =
    \mathcal{O}\left(
        \frac{\tilde{\alpha}_{\mathrm{comm}}^{1/p} t^{1+1/p}}{\epsilon}
    \right) ,
\end{gather}
where $\tilde{\alpha}_{\mathrm{comm}}$ denotes the norm of the commutator, defined as follows:
\begin{gather}
    \tilde{\alpha}_{\mathrm{comm}}
    \equiv
    \left\|
        [H_{\gamma_{p+1}}, \cdots, [H_{\gamma_2}, H_{\gamma_1}] \cdots]
    \right\| .
\end{gather}
When $p$ is sufficiently large, the Trotter number $r$ scales as follows:
\begin{gather}
    r 
    =
    \mathcal{O}\left(
        \frac{\tilde{\alpha}_{\mathrm{comm}}^{o(1)} t^{1+o(1)}}{\epsilon}
    \right) .
\end{gather}

\subsection{Quantum circuit for the controlled two-qubit unitary method}
\label{sec_sm:qc_for_rte_two_qubit}
\begin{figure}[h]
    \centering
    \includegraphics[width=0.8 \textwidth]{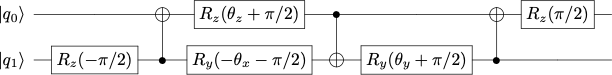} 
\caption{
Quantum circuit for the diagonal two-qubit rotation gate ignores the global phase, $C_{d}(\vec{\theta}) = e^{-i\pi/4} R_d(-\vec{\theta}/2)$.
}
\label{circuit:C_d}
\end{figure}

In Section \ref{sec_sm:qc_for_rte_trotter}, the RTE for the Heisenberg Hamiltonian is divided into the product of two-qubit unitaries. 
PITE and QPE comprise CRTE operations. This section presents a quantum circuit for a controlled two-qubit unitary. 
First, any two-qubit unitary can be decomposed as follows:  
\begin{gather}
    U_{ij}
    =
    e^{i\theta_g } (V_i \otimes V_j)
    R_d(\vec{\theta})
    (W_i \otimes W_j) ,
\end{gather}
where $R_d(\vec{\theta})$ denotes a diagonal two-qubit unitary.
\begin{gather}
    R_d(\vec{\theta})
    =
    e^{
        i(
            \theta_{x} \sigma^{x}_i \otimes \sigma^{x}_j
            +
            \theta_{y} \sigma^{y}_i \otimes \sigma^{y}_j
            +
            \theta_{z} \sigma^{z}_i \otimes \sigma^{z}_j
        )
    } .
\end{gather}
The rotation angles $\vec{\theta} = (\theta_{x}, \theta_{y}, \theta_{z})$, global phase $\theta_{g}$, and single-qubit gates $\{V_{i}, V_{j}, W_{i}, W_{j}\}$ are determined using Cartan's KAK decomposition \cite{Tucci2005arXiv}, which is implemented in Cirq \cite{Cirq}. 
The diagonal two-qubit unitary, $R_d(\vec{\theta})$, is implemented using three CNOT gates, ignoring the global phase denoted by $C_{d}(\vec{\theta})$. 
\begin{gather}
    R_d(\vec{\theta})
    =
    e^{i\pi/4} C_{d}(-2\vec{\theta}).
\end{gather}
 The quantum circuit for $C_{d}(\vec{\theta})$ is depicted in Fig. \ref{circuit:C_d}.

\begin{figure}[h]
    \centering
    \includegraphics[width=0.8 \textwidth]{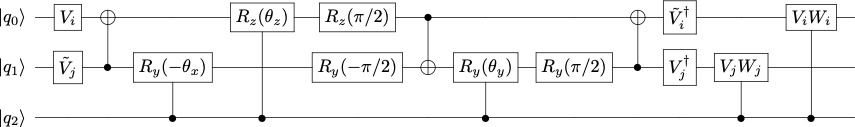} 
\caption{
Quantum circuit for the controlled two-qubit unitary gate using KAK decomposition, ignoring the global phase. The lowest qubit is the control qubit and the other qubits are the target qubits.
$\tilde{V}_j = R_z(-\pi/2)V_j$ and $\tilde{V}_i^{\dagger} = V_i^{\dagger} R_z(\pi/2)$ are used in this figure.
}
\label{circuit:controlled_kak}
\end{figure}

This implementation scheme is the most efficient method for any two-qubit unitary and does not require any approximation, such as Trotter decomposition; thus, we extend the circuit to the controlled two-qubit unitary.
The controlled operation of $U_{ij}$ is realized by enabling the rotation angles $\{\theta_{x}, \theta_{y}, \theta_{z}\}$ when the ancilla qubit is in the $|1\rangle$ state.
In addition, the single-qubit gates, $V_{i}$ and $W_{i}$, do not cancel each other out, and $V_{i} W_{i} \neq I_{2}$. 
Thus, the application of these single-qubit gates depends on the state of the control qubit.
The $W_{i}$ gate is rewritten as $W_{i} = W_{i} V_i V_i^{\dagger}$ to reduce the number of CNOT gates. Then, the $V_i^{\dagger}$ gate is implemented as a single-qubit gate, and the $W_{i} V_i$ gate is implemented as a controlled unitary gate.
The quantum circuit for the controlled unitary, $U_{ij}$, is illustrated in Fig. \ref{circuit:controlled_kak}. 
The following equations demonstrate that the quantum circuit depicted in Fig. \ref{circuit:controlled_kak} is controlled by the $U_{ij}$ operation.
\begin{gather}
    Z_{\theta_g} \otimes I_{4}
    \left(
        |0\rangle \langle 0| 
        \otimes 
        (V_i \otimes V_{j}) C_d(\vec{\theta}=\vec{0} ) (V_i^{\dagger} \otimes V_{j}^{\dagger})
        +
        |1\rangle \langle 1| 
        \otimes 
        (V_i \otimes V_{j}) C_d(-2\vec{\theta}) (W_i \otimes W_{j})
    \right)
    \nonumber \\
    =
    e^{-i\pi/4} Z_{\theta_g} \otimes I_{4}
    \left(
        |0\rangle \langle 0| 
        \otimes 
        I_{4}
        +
        |1\rangle \langle 1| 
        \otimes 
        (V_i \otimes V_{j}) R_d(\vec{\theta})(W_i \otimes W_{j})
    \right)
    \nonumber \\
    =
    e^{-i\pi/4} Z_{\theta_g} \otimes I_{4}
    \left(
        |0\rangle \langle 0| 
        \otimes 
        I_{4}
        +
        |1\rangle \langle 1| 
        \otimes 
        e^{-i\theta_g } U_{ij}
    \right)
    \nonumber \\
    =
    e^{-i\pi/4} 
    \left(
        |0\rangle \langle 0| 
        \otimes 
        I_{4}
        +
        |1\rangle \langle 1| 
        \otimes 
        U_{ij}
    \right) .
\end{gather}

Because the controlled unitary gate is implemented using two CNOT gates \cite{Nielsen2000Book}, the controlled unitary, $U_{ij}$, requires 13 CNOT gates.

\section{Details of numerical results}
\label{sec_sm:results}
\subsection{Setup}
\label{sec_sm:results_setup}
This section presents additional numerical results that provide deep insight into PITE, multi-step PITE, and QPE.
We use the same computational model as in the main text, i.e., the Heisenberg model, expressed as follows:
\begin{gather}
    \mathcal{H}
    =
    \sum_{\langle j,k \rangle} \vec{\sigma}_{j} \cdot \vec{\sigma}_{k}
    +
    \sum_{j} h_j \sigma_{j}^{z} ,
\end{gather}
where $\vec{\sigma}_{j} = (\sigma_{j}^{x}, \sigma_{j}^{y}, \sigma_{j}^{z})$ and $\langle j, k \rangle$ represent the combination of the nearest neighbors of the closed one-dimensional chain. 
$h_j$ represents the strength of the magnetic field, which is randomly selected from a uniform distribution: $h_j \in [-1, 1]$.

As discussed in our previous study \cite{Nishi2023arXiv}, an appropriate constant-energy shift prevents the exponential decay of the total success probability with the progression of PITE steps. 
This constant energy shift is adopted in the numerical simulation. 
In addition, the imaginary-time step size can be varied step-by-step. 
\begin{gather}
    \Delta \tau_k
    =
        (1 - e^{-(k-1)/\kappa})
        (\Delta \tau_{\mathrm{max}} - \Delta \tau_{\mathrm{min}})
        +
        \Delta \tau_{\mathrm{min}}
        ,
\label{eq:exponential_schedule}
\end{gather}
where $\kappa$ is introduced to control the rate of change.
Larger values of $\kappa$ correspond to a rapid change of $\Delta\tau_{k}$ from $\Delta\tau_{\min}$ to $\Delta\tau_{\max}$; smaller $\kappa$ values lead to the opposite case. 
The behavior of the infidelity of PITE was also studied in Ref. \cite{Nishi2023arXiv} by varying the parameters $\Delta \tau_{\mathrm{min}}$, $\Delta \tau_{\mathrm{max}}$, $\bar{\kappa}$.
The authors concluded that incorporating a certain level of nonlinearity into linear scheduling is an effective approach to suppress infidelity and accelerate the computation process involved in ground state preparation. 
Based on previous reports, we set $\Delta \tau_{\mathrm{min}} = \pi / (2s\Delta\tau_{N})$, $\Delta \tau_{\mathrm{max}} = \pi / (s \Delta\tau_{2})$, and $\bar{\kappa} \equiv \kappa / K = 1$.

\subsection{Dependence on the order of the Trotter decomposition for CRTE}
\begin{figure}[h]
    \centering
    \includegraphics[width=0.95 \textwidth]{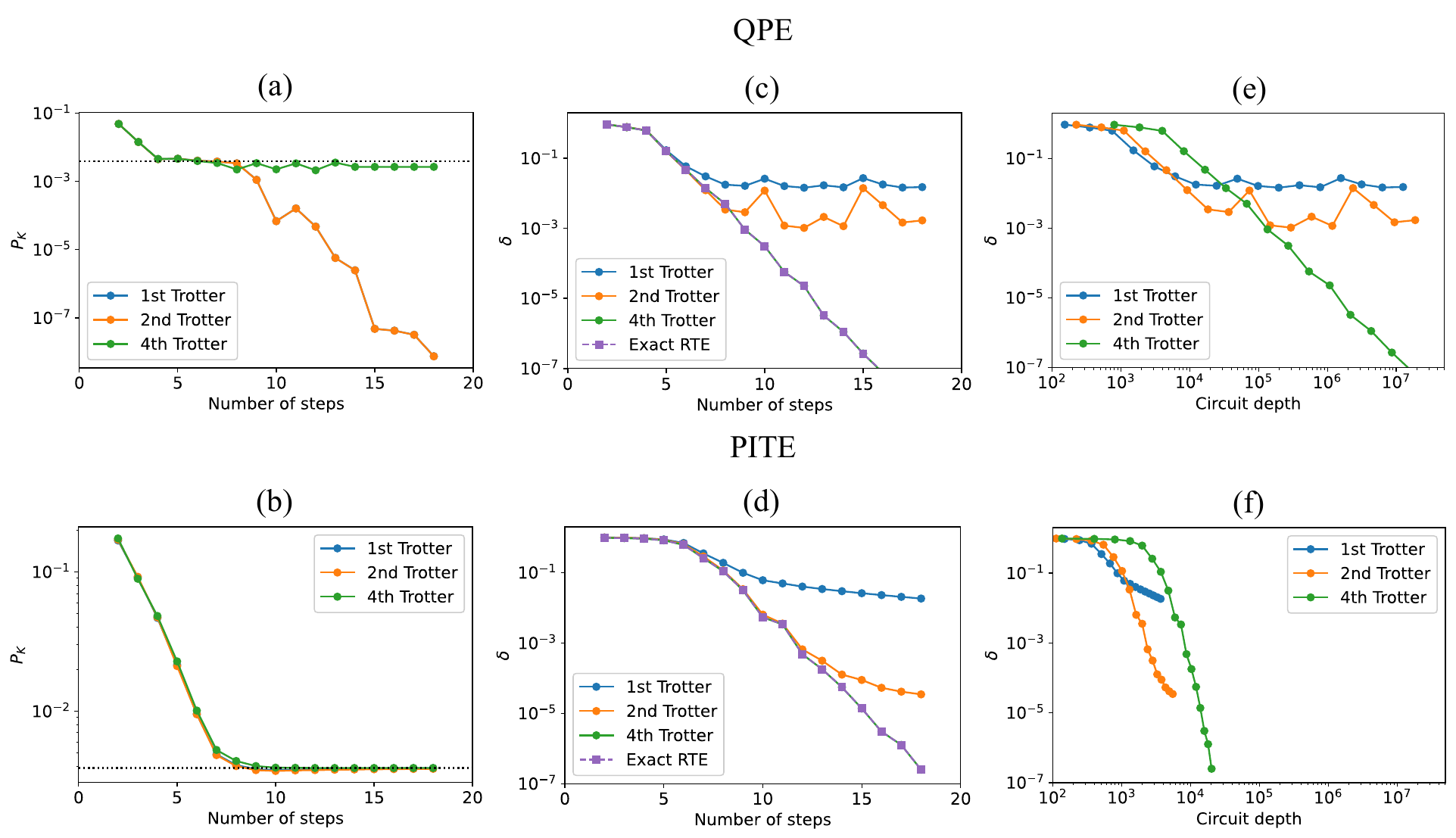} 
\caption{
The numerical simulation for ground-state preparation is performed using the Heisenberg model in a one-dimensional chain with eight spins.
Total success probability, $P_K$, is depicted with respect to the number of steps for (a) QPE and (b) PITE. The dotted black line represents the probability weight of the ground state, which is a theoretically estimated value (see main text). The infidelity $\delta \equiv 1 - \mathcal{F}$ with fidelity $\mathcal{F}$ is plotted by varying the number of steps for (c) QPE and (d) PITE. 
The infidelity is presented as a function of the circuit depth for (e) QPE and (f) PITE. 
The CRTE gates are implemented using first- (blue line), second- (orange line), and fourth-order (green line) Trotter--Suzuki decompositions. The exact implementations of the CRTE gates are indicated by purple lines. 
}
\label{fig:sm_results_pite}
\end{figure}

QPE and PITE are executed by varying the number of steps, $K$, and the total success probability, $P_K$, and the infidelity, defined as $\delta \equiv 1 - \mathcal{F}$ with fidelity $\mathcal{F}$, are plotted in Fig. \ref{fig:sm_results_pite}.
The initial state is prepared as a uniform probability weight for each eigenstate, $|c_{i}| = 1/\sqrt{N}$.
The number of steps for QPE corresponds to the number of ancillary qubits used to determine the accuracy of the eigenvalue in the binary representation loaded into the ancilla qubit.
The CRTE gates in QPE and PITE are implemented using Trotter--Suzuki decompositions; Trotter--Suzuki decompositions of the first, second, and fourth order are used. 
As QPE retains a numerical error due to the CRTE gates, the time of the RTE operator is further divided by $r=4$. 
The scaling parameter, $t_0$, of QPE used to enlarge or reduce the search range of eigenvalues for precise estimation is taken to be $t_0 = 2^{K-N_C}$, where $N_C = \lfloor \log_2(\lambda_N - \lambda_1) \rfloor$. 
Notably, $\{\Delta\tau_{k}\}$ of PITE is calculated using Eq. (\ref{eq:exponential_schedule}), with constant $\Delta\tau_{\max}$, $\Delta\tau_{\min}$, and the number of steps $K$. In other words, $\{\Delta\tau_{k}\}$ is calculated for each plot point.

First, we consider the total success probability, $P_{K}$, in Figs. \ref{fig:sm_results_pite}(a, b). 
As discussed in the main text, the total success probability is given by $P_{K} = |c_{1}|^{2} / (1-\delta) $, which is depicted by the dotted black lines with $\delta=0$. 
The QPE with the fourth-order Trotter decomposition exhibits good agreement with the $P_{K} = |c_1|^{2}$ line. 
In comparison, the first- and second-order cases deviate from the ideal values, and the degree of deviation is prominent for a large number of steps. 
This deviation is considered to induce error due to insufficient Trotter--Suzuki decomposition of the non-commutator terms.
The total success probabilities for PITE converge to the ideal value in small steps for every order of the Trotter--Suzuki decomposition.

The results depicted in Figure \ref{fig:sm_results_pite}(c, d) indicate that the fourth-order Trotter decomposition exhibits good accuracy, coinciding with the exact implementation of the CRTE operator for QPE and PITE. 
The first- and second-order Trotter decompositions deviate from the exact RTE result and saturate the error within a certain infidelity.   
PITE exhibits the same behavior; however, the amount of error decreases as the order of the Trotter decomposition increases. 
PITE is observed to be robust with respect to errors induced by the Trotter–-Suzuki decomposition.

Finally, the circuit depths for the QPE and PITE are estimated and plotted in Figs. \ref{fig:sm_results_pite}(e, f). 
For simplicity, the computational cost of the quantum circuit in the initial state is ignored.
CRTE gates are implemented using Qiskit \cite{Qiskit} and the circuit depth for one CRTE gate, $d_{\mathrm{CRTE}}$, is calculated. 
Subsequently, the circuit depth is estimated based on the depth of the CRTE gate and the number of queries transmitted to the CRTE gates.
The circuit depth for QPE is estimated as follows: 
\begin{gather}
    d_{\mathrm{QPE}}
    =
    1 
    +
    (2^K - 1) r d_{\mathrm{CRTE}}
    +
    d_{\mathrm{QFT}}^{(K)}
\end{gather}
where $d_{\mathrm{QFT}}^{(K)}$ represents the circuit depth of QFT with $K$ qubits.
In Fig. \ref{fig:sm_results_pite}(e), QPE exhibits Heisenberg scaling, where the circuit depth increases linearly as the infidelity decreases.  
In contrast, the circuit depth for PITE increases logarithmically as the infidelity decreases. 
In other words, the analytical estimation presented in the main text is confirmed.
This exponential improvement over QPE arises from the linear increase in the imaginary time-step size, $\Delta\tau_{k}$. 
In fact, exponential scheduling is used; however, the speed of increase of $\Delta\tau_{k}$ is adjusted to be linear using the parameter, $\bar{\kappa}$.
The circuit depth for PITE is calculated as follows: 
\begin{gather}
    d_{\mathrm{PITE}}
    =
    4 
    +
    K(K-1)d_{\mathrm{CRTE}} .
\end{gather}
Accordingly, the number of queries transmitted to CRTE gates increases linearly in the case of PITE, leading to an exponential advantage over QPE with respect to infidelity.

\subsection{Dependence of the ground state in the initial state on probability weight}

\begin{figure}[h]
    \centering
    \includegraphics[width=0.95 \textwidth]{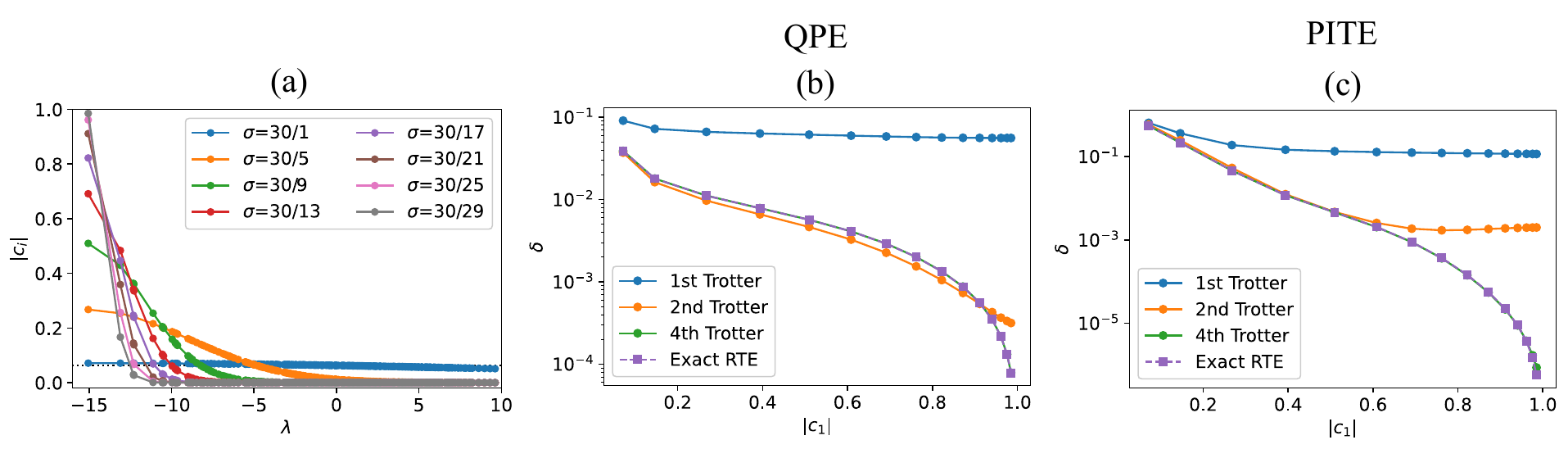} 
\caption{
Numerical simulation for ground state preparation is performed using the Heisenberg model in a one-dimensional chain with eight spins.
(a) Probability weights of the initial state as functions of eigenvalues that follow a Gaussian distribution given by Eq. (\ref{eq_sum:prob_weight_gauss}).
Variances $\sigma$ are plotted in different colors.
The dotted black line represents the uniform probability weight, $|c_i|=1/\sqrt{N}$.
The infidelity is plotted with respect to the ground-state weight, $|c_1|$, for (b) QPE and (c) PITE.  
The number of steps is fixed at $K=6$.
}
\label{fig:sm_results_weight}
\end{figure}

The first numerical simulation is performed using a superposition state for each eigenstate with a uniform probability weight. 
However, this scenario is unrealistic.
Numerous applications have been proposed for the approximate calculation of the ground state using classical computers. 
For example, density functional theory (DFT) \cite{Hohenberg1964, Kohn1965} is widely used in materials sciences.
In addition, the coupled cluster method starts from the initial state obtained via Hartree--Fock (HF) calculation \cite{Bartlett2007RMP, Helgaker2013Book}, and the use of DFT or HF states as an initial state is an adequate scenario in this case.

Judicious selection of the initial state increases the total success probability for QPE and PITE. 
Therefore, we calculate the errors for QPE and PITE with respect to varying probability weights of the initial state.
For example, the initial state weight can be obtained using a Gaussian distribution. 
\begin{gather}
    c_i
    \propto 
    \exp \left(
        -\frac{(\lambda_{i} - \lambda_{1})^{2}}{2\sigma^{2}} 
    \right),
\label{eq_sum:prob_weight_gauss}
\end{gather}
where $\sigma$ denotes variance.
The initial state probability weights used in this study are depicted in Fig. \ref{fig:sm_results_weight}(a). 
The mean of $\lambda_{1}$ becomes concentrated in the probability weight of the ground state by decreasing $\sigma$. 
A uniform probability weight corresponds to a large value limit for $\sigma$.
In the numerical simulation, we set $\sigma = 30/i$ for $i=1, 3, \ldots, 29$.

The infidelities of QPE and PITE corresponding to $|c_1|$ are depicted in Figures \ref{fig:sm_results_weight}(b, c).
The number of steps is fixed at $K=6$. 
In both methods, the infidelity decreases as $|c_1|$ increases. 
An increase in the probability weight of the ground state corresponds to a decrease in infidelity.
In addition, reducing the probability weight of high-energy states decreases the effect of the non-commutator terms in the Trotter--Suzuki approximation.

\subsection{Estimation of circuit depth for multi-step PITE}
The entire quantum circuit of multi-step PITE consists of first-acting PITE and $m^{*}$ iterations of the amplitude amplification operator. The circuit depth of multi-step PITE is given by
\begin{gather}
    d_{\mathrm{PITE+AA}}
    =
    m^{*}
    \left(
        2d_{\mathrm{PITE}} + d_{S_0}^{(K)} + d_{S_0}^{(n+K)}
        +
        2d_{U_{\mathrm{ref}}}
    \right)
    +
    d_{\mathrm{PITE}} ,
\end{gather}
where 
\begin{gather}
    m^{*}
    =
    \left \lfloor
        \frac{(2n+1)\pi}{4\sin^{-1}\sqrt{P_K}}
    \right\rfloor ,
\end{gather}
where $n$ denotes an integer.
$d_{\mathrm{PITE}}$ represents the circuit depth of the approximated PITE over $K$ steps. $d_{S_0}^{(n)}$ represents the circuit depth for zero reflections of $n$ qubits. 
$d_{U_{\mathrm{ref}}}$ represents the circuit depth for the quantum circuit, $U_{\mathrm{ref}}$, which creates the initial state $|\psi\rangle$ from the $|0\rangle^{\otimes n}$ state.

We are particularly interested in the following question: 
When is adopting QAA more beneficial than adopting PITE without QAA?
If $P_K$ is sufficiently large, adopting QAA is inefficient in terms of the computational cost. 
The total success probability, $P_K$, for $m^{*}$ repetitions is expressed as follows:
\begin{gather}
    \sin\left(
        \frac{\pi}{4(m^{*}+1)}
    \right)
    <
    \sqrt{P_K}
    \leq 
    \sin\left(
        \frac{\pi}{4m^{*}}
    \right) ,
\end{gather}
where the upper bound of $\sqrt{P_K}$ is one when $m^{*}=0$.
Tuning the PITE parameter, $\gamma$, leads to the deterministic operation of the ITE operator by increasing the QAA repetition to $m^{*}+1$. 
If $1/2 < P_K \leq 1$ when $m^{*}=1$, it is easy to prove that $d_{\mathrm{PITE+AA}}$ is always higher than the computational cost of PITE.
In other cases, we approximate $m^{*}$ as $m^* \approx \pi/(4\sqrt{P_K})$, indicating that deriving the relationship in the adoption of QAA is beneficial:
\begin{gather}
    d_{\mathrm{PITE}}
    \geq
    \frac{\pi|c_1|}{4(1 - |c_1|^{2})} 
    \left(
        d_{S_0}^{(K)} + d_{S_0}^{(n+K)}
        +
        2d_{U_{\mathrm{ref}}}
    \right) .
\end{gather}
This relation implies that adopting QAA is useful when the implementation cost of zero reflection is lower than that of PITE or when the probability weight of the ground state is sufficiently small.

\bibliographystyle{apsrev4-2}
\bibliography{ref}